\documentclass[aps,prl,showpacs,twocolumn,floats,epsfig]{revtex4}
\usepackage{amssymb}
\usepackage{amsbsy}
\usepackage{amsmath}
\usepackage{epsfig}
\begin{document}
\def\be{\begin{equation}}
\def\ee{\end{equation}}
\def\bearr{\begin{eqnarray}}
\def\eearr{\end{eqnarray}}
\def\tc{$T_c~$}
\def\tcl{$T_c^{1*}~$}
\def\c2{ CuO$_2~$}
\def\ruo{ RuO$_2~$}
\def\lsco{LSCO~}
\def\bi{bI-2201~}
\def\tl{Tl-2201~}
\def\hg{Hg-1201~}
\def\sro{$Sr_2 Ru O_4$~}
\def\rc{$RuSr_2Gd Cu_2 O_8$~}
\def\mgb{$MgB_2$~}
\def\pz{$p_z$~}
\def\ppi{$p\pi$~}
\def\sqo{$S(q,\omega)$~}
\def\tperp{$t_{\perp}$~}
\def\he4{${\rm {}^4He}$~}
\def\ags{${\rm Ag_5 Pb_2O_6}$~}
\def\nxcob{$\rm{Na_x CoO_2.yH_2O}$~}
\def\lsco{$\rm{La_{2-x}Sr_xCuO_4}$~}
\def\lco{$\rm{La_2CuO_4}$~}
\def\lbco{$\rm{La_{2-x}Ba_x CuO_4}$~}
\def\half{$\frac{1}{2}$~}
\def\tst{${\rm T^*$~}}
\def\tch{${\rm T_{ch}$~}}
\def\jeff{${\rm J_{eff}$~}}
\def\nbc{${\rm LuNi_2B_2C}$~}
\def\cabc{${\rm CaB_2C_2}$~}
\def\nboo{${\rm NbO_2}$~}
\def\voo{${\rm VO_2}$~}
\def\h2o{${\rm H_2 O}$~}
\def\nh3{${\rm N H_3}$~}
\def\K3C60{$\rm K_3C_{60}$~}
\def\K3X{$\rm K_3X$~}
\def\K3Pi{$\rm K_3(Picene)$~}
\def\K3pTp{$\rm K_3(pTerphenyl)$~}

\title{Resonating Valence Bond Theory of Superconductivity:
Beyond Cuprates \footnote{Based on a Plenary Talk given at
MSM17 at Sharif University of
Technology, Tehran (18-21 September 2017)}}

\author{G. Baskaran\\
The Institute of Mathematical Sciences, Chennai 600 113, India \& \\
Perimeter Institute for Theoretical Physics, Waterloo, Ontario, Canada N2L 2Y5}

\begin{abstract}
Resonating valence bond (RVB) theory of high Tc superconductivity, an electron correlation based mechanism, began as an insightful response by Anderson, to Bednorz and Muller's discovery of high Tc superconductivity in cuprates in late 1986. Shortly a theoretical framework for quantum spin liquids and superconductivity was developed. This theory adresses a formidable strong coupling quantum manybody problem, in modern times. It is built on certain key experimental facts: i) survival of a dynamical Mott localization in a metallic state, ii) proliferation of bond singlets and iii) absence of fermi liquid quasi particles. After summarising RVB theory I will provide an aerial view of, mostly, new superconductors where I believe that, to a large degree RVB mechanism is at work and indicate prospects for even higher Tc's. 
\end{abstract}

\maketitle

\centerline{{\bf {\large 1. Introduction}}}
~~\\

Bednorz and Muller's discovery of high Tc superconductivity in cuprates \cite{BednorzMuller} marks a turning point in the history of superconductivity. The parent compound, \lco , is a layered perovskite. Electronic, structural complexity of the cuprate family and scale of Tc's posed a challenge in finding the correct mechanism. Resonating valence bond (RVB) theory of high Tc superconductivity \cite{PWAScience} made the turning point even more dramatic, by pointing out inadequacy of the phonon mediated BCS mechanism of superconductivity. It provided a viable alternative by bringing in, in a most natural fashion, \textit{Mott localization and singlet chemical bonds into the realm of metal physics and superconductivity}. 

Right at the turning point, in about a month of Bednorz Muller's discovery, conceptualization and model building was done by Anderson, as described in his paper \cite{PWAScience} in early 1987. His inspiring presence and a dynamic young group at Princeton resulted in about eight papers \cite{BZA,ABZH,BAU1,ZA,SU2,WHA,PiFlux,Liang}, also in a short period. This exciting early period in RVB theory also witnessed a spontaneous, wider and wonderful participation \cite{Kotliar,Ruckenstein,KRS,ZhangRice,Sutherland,Fukuyama,Maekawa,Doniach,Elbio,
Kalmeyer,WiegmannGauge,Dzhyaloshinskii,ZhangGrossRiceShiba,Gros,tkLee,YokoyamaShiba,
MatsuiGauge,MullerHartmann,GBYuLuTosatti}.

In RVB theory, challenges from a strong coupling nature of the problem were circumvented to a large extent, by effective use of basic solid state physics concepts and quantum chemistry. Qualitative success obtained by physically motivated approximate many body theories and conceptual richness that RVB theory has achieved has few parallels in the history of condensed matter physics \cite{PWABook,WenLee,PlainVanilla,GBIran,PWAPersonalHistory,GBMoreIsDifferent,GBRandomWalk,5FoldWay}.

Two dimensional tJ model is the minimal model for the conducting CuO$_2$ planes, in RVB theory. A projective nature, basic to tJ model, has not enabled accurate quantitative calculation of physical quantities, in comparison to fermi liquid based BCS theory. However, there is mounting evidence, accumulated over last three decades, that RVB theory has i) identified the correct minimal model, ii) provided the correct mechanism, namely a strong pairing via singlet bonds assisted by antiferromagnetic superexchange and iii) provided a firm conceptual frame work based on quantitative manybody theory.

Focus of the present article is to gather support for RVB theory from new superconducting materials that have appeared in the scene, after cuprates; and some that were already around. We start with a short introduction to RVB theory for cuprates. This will be followed by a brief discussion of several systems where, in my opinion,  RVB mechanism of superconductivity seems to have a significant contribution. Along the way I will present an optimistic picture that within solid state and quantum chemical constraints, RVB theory allows for superconductivity reaching room temperature scales. Stage seems ripe for a dialogue between theorists and experimentalists to achieve this goal.
~\\

\centerline{\bf{ {\large 2. RVB Theory for Cuprates}}}
\begin{flushleft}
{\bf{2a) Anderson's Solution and a Minimal Model}}
\end{flushleft}

Bednorz-Muller's discovery of superconductivity in doped \lco with an observed $T_c$ ($\sim 35$ K in LSCO) put the Tc near the limit set by the phonon mechanism. The fact that the parent compound \lco is a Mott insulator and discovery of YBCO, that followed heels, with significant higher Tc $\sim$ 90 K,  was compelling enough to find an alternative \cite{PWAScience} to phonon based BCS mechanism. The alternative, RVB theory, is based on the following hypotheses:
~~\\

i) \textit{Mott insulator, rather than a fermi liquid, is the right framework to discuss low energy physics of lightly doped Mott insulators.}
~~\\

ii) \textit{Mott localization continues in the lightly doped metals. As a consequence, local moments, antiferromagnetic superexchange and bond singlets survive, in a dynamical fashion, in the metallic state.}
~~\\

iii) \textit{Single band tJ model, a minimal model, captures the above physics, via projection
and superexchange J}.
~~\\

The minimal model for the parent Mott insulator \lco is the spin-half Heisenberg antiferromagnetic Hamiltonian on a square lattice. It describes spin dynamics, at low energies, well below the Mott Hubbard charge gap:
\be
H_{m} = + J \sum_{\langle ij \rangle} ({\bf S}_i \cdot {\bf S}_j - \frac{1}{4})
\ee
Here J $\sim$ 150 meV, is the antiferromagnetic superexchange between two neighboring cites. Spin half moment resides in a wave function that is a linear combination of Cu 3d$_{x^2-y^2}$ atomic orbital and nearest oxygen 2p orbitals. This hybrid has d$_{x^2-y^2}$ symmetry. 

From small to optimal doping regime, a minimal model for the correlated metallic state is the tJ model:
\bearr
H_{\rm tJ} = - t \sum_{\langle ij \rangle} ( c^\dagger_{i\sigma}c^{}_{j\sigma}  + h.c.)  + J \sum_{\langle ij \rangle} ({\bf S}_i \cdot {\bf S}_j - \frac{1}{4}n_i n_j)
\eearr

with an important local constraint (for hole doping)  $ n_{i\uparrow} + n_{i\uparrow} \neq$ 2, for every site. The hopping parameter t $\approx$ 250 meV. The local constraint seriously limits independent electron hopping: i.e., an electron can only hope to an empty site, if available. This leads to a \textit{projected Many fermion Hilbert space}, where fermi liquid vacuum and fermi liquid quasi particle excitations are simply absent for small doping. 

Super (kinetic) exchange, a local quantum chemical process, provides a large energy scale for spin pairing and formation of preformed spin singlet pairs. Superexchange is the strong pairing glue, responsible for high scale of superconducting Tc seen in doped cuprates. 

It should be pointed out that the above minimal model can be improved and adapated for various members of the cuprate family, to get quantitatively accurate phase diagram etc., by inclusion of next nearest neighbor hopping t', for example.

Anderson's original proposal \cite{PWAScience} suggested plausible solutions to tJ model, using physical arguments and a class of variational RVB wave functions. Anderson generalised his (1973) RVB theory \cite{PWARVB1973} of geometrically frustrated 2d Heisenberg spin system of Mott insulator, and suggested spin liquid states for undoped and doped cuprates.

Now in 1987, Anderson's RVB variational wave function, for both undoped and (hole) doped Mott insulator, was a BCS wave function in electron Fock space, but with double occupancies projected out by Gutzwiller projection. \textit{Writing the spin liquid state of a Mott insulator and a superconductor, in an unified fashion, as a Gutzwiller projected BCS paired electron state, was a revelation}. It turned out to be a key step in the development of RVB theory of superconductivity: 

\be
|{\rm RVB};\phi\rangle \equiv P_G (\sum_{ij} \phi_{ij} b^\dagger_{ij})^{N_e\over2}|0\rangle.
\ee

Here N$_e$ is the total number of electrons. The singlet pair creation operator 
$b^\dagger_{ij} \equiv \frac{1}{\sqrt{2}} (c^\dagger_{i\uparrow}c^\dagger_{j\downarrow} - c^\dagger_{i\downarrow}c^\dagger_{j\uparrow})$. The reference N$_e$ electron state $(\sum_{ij} \phi_{ij} b^\dagger_{ij})^{N_e\over2}|0\rangle$, is a N$_e$ particle projection of the standard BCS wave function $ \prod_k (u_k + v_k c^\dagger_{k\uparrow}c^\dagger_{-k\downarrow})|0\rangle$. Gutzwiller projection 
$P_G \equiv \prod_i (1- n_{i\uparrow}n_{i\downarrow})$ further acts on this N$_e$ particle BCS wave function and projects out double occupancy at all sites.

Further $\phi_{ij}$ is a \textit{Cooper pair or valence bond function}; it is the Fourier transform of the function $(\frac{u_k}{v_k})$. For the popular short range RVB state \cite{KRS,Sutherland} $\phi_{ij} = \phi_0 \neq 0$, only when sites i,j are nearest neighbors. At half filling number of electrons N$_e$ = N, the number of lattice sites; in this Mott insulator we have a neutral quantum spin liquid. When N$_e$ = N(1-x), we have a hole doped Mott insulator in a superconducting state with a doping fraction of x. 

In Anderson's proposal, i) the ground state of the parent Mott insulator is (close to) a quantum spin liquid, a quantum liquid of neutral resonating singlet bonds, with a finite gap (Mott Hubbard gap) for charge excitations, ii) these preformed Cooper pairs represent only fluctuations of spins in a \textit{charge incompressible} quantum spin liquid and iii) on doping, neutral spin singlet pairs get charged, carry current and \textit{a Mott insulator metamorphise into a superconductor}. Strength of superconductivity (superfluid stiffness in the CuO$_2$ planes) is proportional to doping x, for small x. 

Long range antiferromagnetic order seen at zero and very low doping in cuprates is only a mild perturbation in an otherwise robust reference spin liquid state \cite{Hsu}. Finite temperture and doping induced career dynamics quickly melts long range magnetic order and exposes a robust spin liquid state.

~~\\
\begin{flushleft}
{\bf{2b) RVB Theory and Emergent Gauge Fields}}
\end{flushleft}
Having identified a minimal model, we proceeded to solve the model as follows. At a technical level, we used a Fierz identity due to Noga \cite{Noga}, of rewriting the antiferromagnetic exchange coupling (second term in equation 2) interms of bond singlet Cooper pair number operator 

\be
J \sum_{\langle ij \rangle} ({\bf S}_i \cdot {\bf S}_j - \frac{1}{4}n_i n_j)  = -J\sum_{\langle ij \rangle} b^\dagger_{ij}b^{}_{ij}
\ee

The fact that superexchange is the pairing glue is manifest in the above form; further the form suggested itself, in the light of Anderson's RVB prposal, a BCS-Bogoliubov meanfield factorization.

RVB mean field theory \cite{BZA}, is a first step in the analysis of the tJ model: we approximate the local constraint by a global one and replace hopping matrix element t by xt (doping x is the probability of finding a hole for an electron to hop). Mean field solutions thus obtained is used as an input in the RVB vatiational wave function (equation 3). Gutzwiller projection, the next step, extracts key non perturbative physics in the ground and low energy excitations of the RVB State quantitatively. 

For example, Gutzwiller projected mean field Bogoliubov quasi particles for Mott insulators become neutral spinon excitations of the spin liquid. The first RVB mean field solution \cite{BZA} for the square lattice, the \textit{zero flux} solution, discovered a U(1) spin liquid containing a pseudo fermi surface for spinon excitations. The next one was the \textit{$\pi$-flux} solution of Affleck and Marson, who discovered \cite{PiFlux} spinons with Dirac nodes.

Renormalization contained in RVB mean field theory has been improved using \textit{Gutzwiller appromiation} \cite{ZhangGrossRiceShiba,PlainVanilla} and slave particle formalism \cite{Kotliar}. Several quantitative comparisons have been made with experiments, including behavior of superconducting gap as a function of doping and behavior of electron spectral functions \cite{PlainVanilla,Nandini,YRZ}.

RVB meanfield analysis readily gives extended-S wave \cite{BZA} and d$_{x^2-y^2}$ wave symmetry \cite{Kotliar} solutions for the spin singlet superconducting order parameter. Theory and experiments have confirmed validity of d-wave solution, containing nodal quasi particles for cuprates.

Using a functional integral approach, which went beyond mean field theory, an emergent U(1) (RVB) gauge field present in spin liquids was discovered \cite{BAU1}. Emergent gauge fields and their sources control low energy physics of the spin liquid. It was soon shown by Wen, Wilczek and Zee \cite{WWZ} and Wiegmann \cite{WiegmannGauge} that emergent RVB U(1) gauge flux measures scalar chirality of spins, ${\bf S}_i\cdot ({\bf S}_j \times {\bf S}_k)$. Sachdev and Read \cite{SachdevRead} showed that electric component of emergent RVB gauge field is related to valence bond order. Further, U(1) gauge symmetry was found to be a subgroup of a larger, SU(2) symmetry \cite{SU2} of the system.

The functional integral approach \cite{BAU1} also found Ginzburg-Landau lattice action for the RVB system, interms of U(1) link variables. At half filling, this microscopic free energy has U(1) local gauge invariance. Elitzur's theorem precludes possibility of spontaneous breaking of local gauge symmetry. It immediately followed that superconductivity is absent in the Mott insulator. In this sense gauge theory approach is a significant step beyond RVB mean field theory.  

Doping appeared as a new term that reduced a local U(1) gauge invariance into a global U(1) invariance. Doping induced superconductivity, a spontaneous breaking of U(1) global symmetry followed. Emergent gauge fields in qauantum spin liquids is continuing to be an active field \cite{WenBook,gaugetheory2,dhLee}. Earlier works \cite{MatsuiGauge,Drzazga} and recent works by Ramakrishnan and collaborators \cite{RVBGLRamakrishnan} have made quantitative progress using RVB Ginzburg Landau action for cuprates.

A carefuly study of RVB variational wave function for tJ model, with input from RVB mean field theory, has given us good insights into the mechanism and ground state properties of high Tc cuprate superconductors in the superconducting phase. Various analytical and varity of numerical works that analyse tJ model and large U Hubbard model directly, close to half filling, have brought out, over years, insights and quantitative understanding
\cite{Jarrel,Kotliar2,Tremblay,Spalek,Millis,Sorella,GarnetChan,Aoki,Corboz} of superconducting pairing that supports RVB scenario.

Calculation of electron spectral functions, dynamical sturcture factors, frequency and T dependent conductivity etc., of undoped, pseudo gap phase and optimally doped superconducting and normal states remains to be done. 

In a remarkable fashion emergent gauge field became manifest in a spin liquid phase in the exactly solvable Kitaev spin-half model in the 2d honeycomb lattice \cite{KitaevHoneyComb}. Further, unlike the Heisenberg model on the same lattice, RVB mean field theory becomes exact \cite{NayakBurnel} for Kitaev model. 
~~\\

\centerline{\bf{{\large 3. RVB Superconductors Beyond Cuprates}}}
~\\
In what follows we comment on few superconducting compounds known at the time of Bednorz Muller's work and several new ones, discovered because of an increase in awareness of electron correlation based mechanism in the materials community. In my opinion enough attention has not been paid to see RVB physics in these systems. What I am going describe in the rest of this artilce is mostly my personal attempt in this direction. Part of this is described in my article \cite{5FoldWay} `Five fold way to new superconductors', an attempt at a synthesis and unification of mechanism of superconductivity in seemingly different systems, with electron correlation effects as an unifying theme.

\begin{flushleft}
{\bf 3a) Organic Superconductors as \\
~~~~~Self Doped Mott Insulators}
\end{flushleft}

ET salts \cite{OrganicSCBook}, in my opinion, provide very good example for RVB mechanism of superconductivity. Members of this family, typically, form a well isolated and single half filled band of spin-half Mott insulators. They are quasi 2 dimensional and exhibit AFM order or spin Peierls dimerization and in few cases a quantum spin liquid phase. Under chemical or physical pressure superconducting half dome appears, across a first order phase boundary. A major difference of ET salts from cuprates is that there is no external doping and the single band crossing the fermi level is always half filled. Absence of external doping was a kind of mental block to appreciate RVB physics in organic superconductors for a while. Bechgaard salts \cite{Bechgaard} are quasi one dimensional organic systems, again a half filled band system.

In a Mott insulator electron occupancy at a site (valency) remains at unity and there is \textit{no real valence fluctuation}. However, some \textit{virtual valence fluctuations} are present and generate J, the antiferromagnetic superexchange. I viewed \cite{GBOrganics} metallization as emergence of a small but finite amount of real valence fluctuations, and called it \textit{self doping}. That is, across the insulator to metal transition, the systems spontaneously generates and maintains self-consistently a small and \textit{finite density x of holons and same density of doublons}, to gain delocalization and Madelung energies. Self doping x is determined by pressure, band parameters, dielectric constant, Madelung energy etc. In this correlated metallic state superexchange J locally survives (the way superexchange survives in cuprates after a small external hole doping). 

It is desirable that self doping x and superexchange J are present manifestly in a low energy model of the correlated metallic state at half filling. A two species tJ model \cite{GBOrganics} that explicity has self doping x and superexchange J was introduced by the present author. This model describes dynamics of an equal number of (small density) of sites having holon and doublons, in the background of superexchange in the rest of the singly occupied sites. This approach successfully unified superconductivity in cuprates and in ET salts.. 

The notion of self doped Mott insulator or self doped RVB states that we introduced for organic superconductors plays very important role in some of the later systems we have studied: K$_3$C$_{60}$, K$_3$(picene), K$_3$(pTerphenyl), impurity band Mott insulators in boron doped diamond, Fe arsenide family, pressurized H$_2$S and certain doped Band insulators. 

\begin{flushleft}
{\bf{3b) K$_3$X (X = C$_{60}$, picene, pTerphenyl, ... ) as\\
~~~~~Self Doped Mott Insulators}}
\end{flushleft}

Discovery of superconductivity in K$_3$C$_{60}$ \cite{C60Superconductivity} came as a pleasant surprise
in 1990. Erio Tosatti and I were quick to suggest a theory \cite{C60GBErio}, which in a modern
perspective could be summarised as follows. Nominal valency of K$_3$C$_{60}$ is K$^{3+}C_{60}^{3-}$. 
The molecule C$_{60}$ is a p-$\pi$ bonded molecular RVB system, where intra molecular electron
correlations are important. Consequence a low spin (S = \half) state is stabilized in the ion
C$_{60}^{3-}$ ion. That is, a Hund coupling favored high spin (S = $\frac{3}{2}$) state for 3 added
valence electrons in the 3 fold degenerate LUMO is destabilized. 

A weak overlap of the LUMO orbitals between neighboring C$_{60}$ molecules results in a band width, small compared to Hubbard U of the LUMO orbitals. It results in a spin-half Mott insulator and antiferromagnetic superexchange couplings. Further, this spin system is frustrated, because of fcc lattice and potential orbital degeneracies. In view of Mott localization there is no valence fluctuation in C$_{60}^{3-}$ ions. 

Reality is little more complicated ! Experimentally \cite{C60Superconductivity} stoichiometric K$_3$C$_{60}$ is a narrow band (bad) metal, indicating presence of a small amount of real valence fluctuation, involving C$^{2-}_{60}$ and  C$_{60}^{4-}$. To rationalize this, we introduced a notion of stability of molecular singlets (SMS) arising from electron correlation effects in the p-$\pi$ pool of C$_{60}$, which stabilises a spin singlet rather than S = 1 state, in doubly charged C$_{60}^{2-}$ and quadruply charged C$_{60}^{4-}$ molecular ions. Chakravarty and Kivelson \cite{C60SudipSteve} suggested a similar idea, called it electron correlation induced singlet electron pair (charge 2e) binding. 

Stability of molecular singlet, energy gain via delocalization and overall Madelung energy causes a Mott insulator to correlated metal transition. Consequently solid K$_3$C$_{60}$ maintains self-consistently a small amount of valence fluctuation, by having an equal number of (small density) of sites having C$_{60}^{2-}$ (holon) and C$_{60}^{4-}$ (doublon), in the background of superexchange in the rest of the C$_{60}^{3-}$ sites.  In other words, this is a self doped RVB system, similar to ET salts we discussed above, in a multi band setting; it supports high Tc RVB superconductivity. 

While theories based on phonon mechanism dominated the scene for a while, our proposal was supported 
by a later experimental discovery \cite{Iwasa} that i) in K$_3$C$_{60}$NH$_3$, insertion of 
NH$_3$ molecule expands the lattice (a negative pressure) and converts the metal into a spin-half 
Mott insulating antiferromagnet and ii) a modest positive external pressure recovers 
superconductivity via a Mott insulator to superconductor transition. More recently new Mott 
insulators based on C$_{60}$ molecules \cite{C60MottInsulators} have been synthesized; 
they all become superconducting under pressure, further supporting our theory. 

In the family of aromatic hydrocarbon molecules, superconductivity in K$_3$(picene) \cite{K3Picene} and few other aromatic molecules \cite{K3AromaticHCMoleculesReview} have been discovered. K$_3$(picene) shows weak signals for superconductivity \cite{K3PiceneORNL} at temperatures as high as  35 K. Picene is a less symmetric molecule, formed by fusing 5 benzene rings in a zig-zag fashion. Another interesting molecule is p-Terphenyl, it is obtained by bridging three benzene rings by two C-C bonds, in a linear fashion. Very recently, in a series of works \cite{K3pTeterphenyl}, K$_3$(pTerphenyl) has been shown to exhibit Meissner effects at temperatures as high as 123 K ! There are also interesting experimental works that seem to offer indirect supports \cite{pTerphenylDessau,pTerphenylHHWen}. 

The present author has suggested \cite{GBpTerphenyl} that self doped Mott insulators are responsible for high Tc superconductivity, in a more complex structural setting, because of reduced symmetries of the aromatic molecules. We have further suggested that in K$_3$(pTerphenyl), cation K$_3^{2+}$ is only doubly charged and forms a fermionic Mott insulator subsystem; doubly charged anion X$_3^{2-}$ forms a Bosonic Mott insulator, taking advantage of local RVB p-$\pi$ pool in pTerphenyl molecule. Internal charge transfer bettween two Mott insulators (mutual or internal doping) have been suggested to support superconductivity based on RVB mechanism. The work by Tosatti and collaborators \cite{pTerphenylTosatti} also invokes the correlated metallic character, but invokes certain molecular modes  in addition.

\begin{flushleft}
{\bf 3c)  Self Organized or Emergent Mott Insulators \\
~~~~~~From Band Insulators}
\end{flushleft}

We have recently suggeste that starting from band insulators we can generate \textit{Mott insulating subsystems}. We call them \textit{emergent or self organized Mott insulators}. We have analysed three experimental examples, as representing emergent Mott insulators that also exhibit superonductivity. They are  i) pressurized molecular solid H$_2$S, ii) Boron doped diamond and ii) lightly doped band insulator LaOBiS$_2$ family. We will briefly discuss them in the same order.

\textbf{Pressurized Solid H$_2$S}. In recent reports Eremets et al., report superconductivity \cite{EremetsH2S} in H$_2$S, a hydrogen rich solid, with Tc as high as $\sim$ 203 K, at a pressure $\sim$ 200 GPa. It is believed that H$_2$S molecule dissociates around 60 GPa; it results in phase separation and formation of a new hydrogen rich compound H$_3$S. Even though structure is not known experimentally, there are good indications that (large) sulfur atoms occupy more space and a strong bonding among themselves. It results in a saturated covalent network (for example helical chains in some cases) of sulfur atoms. Neutral H atoms and H$_2$ molecules get accommodated at small interstitial space. 

Atomic radius of sulfur atom, 1 Au is significantly large compared to radius 0.5 Au of H atom. Consequently lattice parameter of the interstial H atoms is large compared to the size of neutral H atoms. Our hypothesis \cite{GBH2S} is that these \textit{liberated interstitial H atoms (containing a single electron), form a relatively narrow band and a Mott insulating subsystem emerges}. Some theoretical crystal structures and band parameters support our hypothesis. There is also propensity for spin singlet formation in the Mott insulating subsystem, due to a large superexchange J, arising via bridging sulfur orbitals.

Electron affinity of sulfur subsystem is in general different from that of the H subsystem. In general it results in a charge transfer between H and S subsystems. This results in internal doping of the Mott insulator and superconductivity based on RVB mechanism.

My estimates of the doped Mott insulator parameters, for a few structures for H$_2$S, available in the literature from LDA calculation suggests possibility \cite{GBH2S} of superconductivity reaching the scale of 200 K, as seen in the experiment. 

Our picture goes through for hydrogen rich solids in general. A recent pressure induced superconductivity in PH$_3$ \cite{EremetsPH3} with Tc exceeding 100 K can be also explained by our theory.

\textbf{Boron doped diamond} exhibits an insulator to metal transition as a function of doping, called Anderson-Mott transition. Interestingly boron doped diamond exhibits superconductivity \cite{BDopedDiamond}, at the insulator to metal transition point. A boron atom that replaces a carbon creates an acceptor impurity state, a hole with a spin half moment. Randomly distributed acceptor states weakly overlap and form a narrow impurity band. Electron correlations produce random antiferromagnetic coupling. As in phosphorus doped Si, this forms a \textit{valence bond glass} phase \cite{BhattLee}. That is, frozen local singlet bonds, with varying strength and size get organized in a glassy fashion. 

I suggested \cite{GBDiamond} that the Anderson-Mott insulator to metal transition, as a function of Boron doping, can be viewed as a pressure induced phase change in a reference impurity band, that continues to be at half filling. Across the insulator to metal transition it becomes a self doped impurity band Mott insulator. Physics is similar to self doped Mott insulator that we discussed earlier, with an added complication of randomness arising from random boron impurity sites. 

This theory explained various experiments satisfactorily. Further, it also opened door for \textit{impurity band superconductivity} (and other phases) \cite{GBIBMI} in the rich world of band insulators. One of our prediction was that if one can dope nitrogen in diamond and reach metallization, the Tc could be as large as 100 K. Unfortunately structural and quantum chemistry has prevented such a high doping and metallization from N doping in diamond so far. Currently popular Nnitrogen vacancy centers in diamond, at sufficiently high doping can cause electron correlation induced high Tc superconductivity !

{\bf Doped Band Insulators LaOBiS$_2$, HfNCl etc}. On another front, under normal pressure, a family of lightly \textit{doped band insulators} show  superconductivity and intriguing physical properties that resemble doped Mott insulating cuprates. LaOBiS$_2$ is a prototype band insulator belonging to this family. In the electron doped LaO$_{1-x}F_xBiS_2$, for x $\sim$ 0.1, a low density of electron carriers are added to the valence band and superconductivity is seen \cite{Missiguchi}. Recently I focussed \cite{GBLaOBiS2} on LaO$_{1-x}F_xBiS_2$ family.

The dilute density of doped carriers in the band insulator, make use of a special orbital hybridization in the 6p valence orbitals of BiS$_2$  and residual long range coulomb interaction and self organize into emergent lower dimensional Mott insulator structures, having strong local bond singlet correlations. This dynamical and fragile Wigner crystali like organization or electron pairs, leads to the possibility of RVB superconductivity, in a doped band insulator \cite{GBLaOBiS2}. 

\begin{flushleft}
\bf{3d) MgB$_2$, Graphene, Silicene, Germanene ...}
\end{flushleft}

MgB$_2$ is formed by stacking alternate layers of honeycomb lattice of B$^{1-}$ ions and triangular lattice of Mg$^{2+}$ ions. This makes Boron layers electronically similar to graphene, as B$^{1-}$ ion has same electron configuration as a carbon atom. Sigma bands of B$^{1-}$ floats up to fermi level and transfer some electrons to the p$-\pi^*$ band. This is an internal electron doping. MgB$_2$ is a type-II superconductor \cite{MgB2Akimitsu} with a Tc close to 40 K.

While developing a model for valence bond based superconductivity for MgB$_2$, in 2002, based on its similarity to graphite I found \cite{GBMgB2} possibility of high Tc superconductivity in doped single sheet graphite, the well known graphene. I introduced a tJ0 model, a tJ model with no local constraints. The rationale behind tJ0 model is as follows. 

Graphite is not a Mott insulator - it is quasi two dimensional semi metal. The intermediate electron
correlations present in the broad $\pi-\pi^*$ bands of graphene layer, is unable to create Mott
localization. However, being two dimensional, it manages to create an enhanced spin singlet (valence
bond) correlations a la' Pauling, in the semimetal, compared to a free electron semi metal. (Jafari
and the present author prediceted \cite{GBAkbar} that as a consequence of enhanced local spin singlet
correlations, an isolated spin-1 collective mode emerges in the window of the
particle-hole continuum of graphene).

Our tJ0 model combined band physics and valence bond (spin singlet) physics in a semi microscopic fashion, leading to interesting insights.  Our work was pursued by Doniach and Black-Shaffer \cite{DoniachAnnica}, who in addition to  confirming very high scale of superconducting Tc's, discovered an unconventional d + id  chiral superconductivity solution. A variational Monte Carlo calculation \cite{PathakShenoyGB} with Vijay Shenoy and Sandeep Pathak, that went beyond mean field theory and took into account quantum fluctuations gave a numerical confirmation of chiral superconductivity. The estimated Tc was high $\sim $ 200 K ! Subsequent theoretical developments have confirmed the above \cite{NandKishoreThomaleZhengcheng}. Experiments have not confirmed our prediction, possibly because of an unavoidable disorder that comes at the desired range of doping and a high sensitivity of d + id state to disorder. That is, even when one achieves the desired optimal doping, disordered induced proliferation of chiral domains could hide a high Tc superconductivity in the experiments.

I should also point out that experimental groups of Kopelevich and Esquinazi \cite{KopelevichEsquinazi} have reported unstable and tenuous signals for superconductivity, reaching room temperature scales in perturbed (doped) graphite. These signals, because of the above mentioned theoritcal possibility, are worth exploring further experimentally.

Silicene, Germanene and Stanene are 2d analogues of graphene. C,Si,Ge and Sn occur in the same column in the periodic table. However the atomic radii of Si and Ge are about 60 percent larger than that of carbon. According to electronic structure calculations this leads to a substantial, 3 fold reduction in the band width of p-$\pi$ band in silicene and germanene.  Based on band theory estimates of the t and U parameter and some overwhelming phenomenology I came to the conclusion that silicene and germanene are likely to be Mott insulators \cite{GBSilicene}. In collaboration with Vidhya \cite{VidhyaGB} we have found, in a state of the art ab-initio calculation, evidence for valence bond ordering instabilities in silicene, essentially a stretched graphene. This result indicates possibility of Mott localization in silicene, in the density functional framework.

The t and J parameter I estimate for silicene makes it a prospective playground for room temperature superconductor after an optimal doping, provided competing phases such as valence bond ordering are kept under control.

\begin{flushleft}
\bf{3e) Nickel Borocarbides}
\end{flushleft}

Nickel borocarbide family of superconductors \cite{NBC} is a layered system. Interestingly the layer structure is similar to the FeAs layers in the Fe based superconductors, discovered much later. I was convinced of a mechanism of superconductivity based on electron correlation \cite{GBNBC}, based on quantum chemistry. Neutron scattering also showed \cite{NBCNeutronResonance} a ($\pi,\pi$) mode at 4 meV, that becomes sharp below Tc,  resembling the neutron resonanace mode in cuprates. In view of a complicated looking band structure chand fermi surfaces, modelling was some what complicated and a deeper understanding remains obscure.

\begin{flushleft}
\bf{3f) Hydrated Sodium Cobalt Oxide}
\end{flushleft}

Na$_x$.CoO$_2$.yH$_2$O, a hydrated sodium cobalt oxide, has CoO$_2$ layers, intercalated with Na ions and two dimensional ice (H$_2$O) sheets. It becomes superconducting at low temperatures \cite{NaCoO}. From existing band structure results, a narrow conduction band and correlation based physics was some what obvious. Further Co atoms formed a triangular lattice- I predicted spin singlet d + id chiral RVB type of superconductivity \cite{GBNaCoO}. Charge ordering in the CO$_2$ layer and ordering in the intercalant Na layer complicates the matter.

It is clear however that 2d ice sheets strongly screen coulomb repulsions among carriers in the CoO$_2$ plans; this prevents a competing charge density order and encourage superconductivity. Our prediction of  unconventional chiral spin singlet superconductivity, if experimentally discovered, will be an important support for RVB mechanism.

\begin{flushleft}
\bf{3g) Fe Arsenide and Fe Chalcogen Family}
\end{flushleft}

After cuprates, Fe arsenide and Fe chalgogenide family of superconductors are the most thoroughly investigated family of high Tc superconductors \cite{FeAs,FeSeMKWu}. Focussing on quantum chemistry and phenomenology I suggested that this family is an example of double RVB system \cite{GBFeAs}. That is, two valence electrons in the 3d$^6$ shell of Fe$^{2+}$ living in two nearly degenerate 3d orbitals form a pair of (weakly) Hund coupled two dimensional spin half Mott insulating RVB system. Internal charge trnsfer (self doping) between them generate RVB superconductivity. Unfortunately this family is complex, unlike the cuprates, because of an apparently complex multi orbital character. Inspite of the multi orbital character in Fe systems, there are striking phenomenological similarities to the single orbital cuprates.

Based on various considerations we suggested that members of Fe Arsenide family are best thought of spin-1 Mott insulator or self doped spin-1 Mott insulators, depending on the internal chemical pressure that the system has. For example, from a parent spin-1 Mott insulator, isostructural LaOFeAs and SmOFeAs, have undergone a insulator to metal transition, via self doping, due to a larger and smaller internal chemical pressures respectively. A high self doping in LaOFeAs gives a smaller sublattice antiferromagnetic moment; a relatively smaller self doping in SmOFeAs gives a larger moment, close to S = 1, in neutron scattering experiments.

\begin{figure}[t]
\epsfxsize 8.5cm
\centerline {\epsfbox{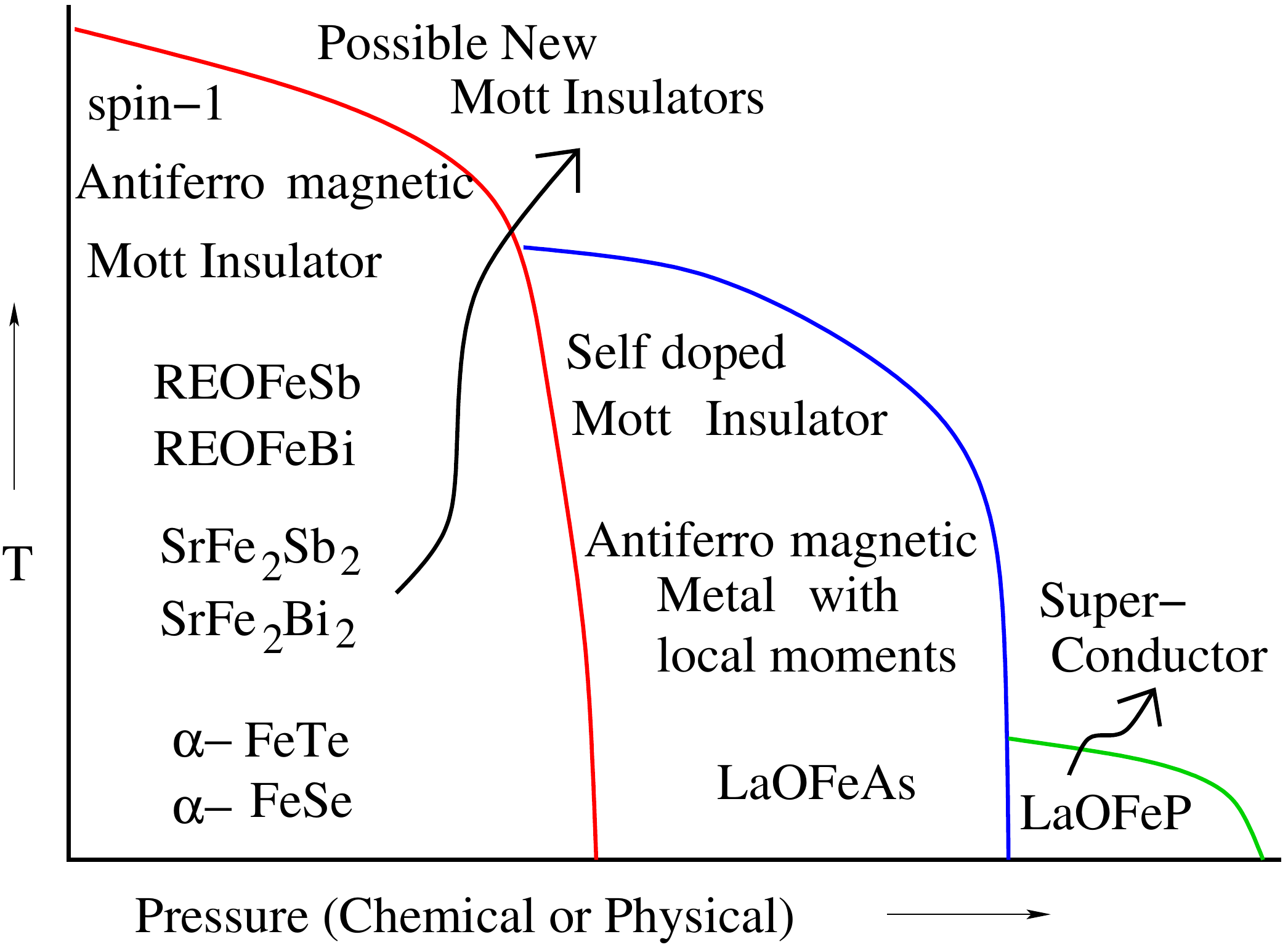}}
\caption{Figure from Reference \cite{GBFeAs}. Schematic temperature vs pressure (chemical or physical) phase diagram. LaOFeAs, a lightly self doped spin-1 Mott insulator is more correlated than the heavily self doped metallic LaOFeP. \textbf{This 
work predicted, in 2008, a spin-1 Mott insulating phase in FeSe}, as arising from decrease in chemical pressure (lattice expansion) and consequent increase in electron correlation effects; \textbf{recent experiments \cite{FeAsMottInsExpt} are in conformity with our prediction}.}
\end{figure}

Based on these considerations we predicted a phase diagram in 2008. It is comforting to see that FeSe, predicted to be S = 1 (intermediate spin) Mott insulators has been confirmed experimentally \cite{FeAsMottInsExpt}. Further, they exhibit maximum superconductivity in the family and supports our picture of doped double RVB superconducting states. 

\begin{flushleft}
\bf{3h) Chiral RVB Superconductivity in\\
~~~~~Doped TiSe$_2$, Spin Ladder Compounds etc.}
\end{flushleft}

In the presence of frustration and special geometry of the lattice, such as triangular or honeycomb lattice, stable chiral RVB mean field solutions that spontaneously break parity and time reversal invariance appear. Two systems where we have theoretically predicted chiral spin singlet supercoductivity are spin ladder compound \cite{GBSpinLadder} and doped TiSe$_2$ \cite{GaneshGB}. This is in line with our earlier work on doped graphene, silicene, germanene and hydrated sodium cobalt oxide.  
~~\\

\centerline{\bf{{\large 4. Hurdles to RVB Mechanism}}}
~~\\
In the case of BCS superconductors magnetic impurities degrade superconducting Tc significantly via pair breaking. The stable singlet bonds in RVB systems are affected in a very different fashion. We find that a major hurdle comes from competing orders, encouraged by doped Mott insulators. Some of the hurdles are: i) strong electron-lattice coupling induced localization (lattice trapping) of singlet bond or Cooper pair via lattice dimerization; a striking example is the competing charge and spin stripe valence bond (pair density wave) orders seen in the low Tc under doped regime in cuprates. ii) Phonons, even without causing localization of singlet bonds or charges, could reduce zero momentum Cooper pair condensate via decoherence of holons and doublons. iii) Another source of strong supression of Tc is presence of a second broad electronic band (a fermi liquid), whose soft particle-hole pair excitations interfere with coherent valence bond delocalization or RVB physics in the first band. MgB$_2$ offers a good example \cite{GBMgB2}, where fermi sea of a broad sigma-band interferes with coherent delocalization of valence bond pairs in the $\pi^*$ band of boron sheets.

\textbf{Competing Hund Coupling and Chiral P Wave Superconductivity in Sr$_2$RuO$_4$} \cite{MaenoSRO}. 
Thanks to Piers Coleman's provocative comments at a strong correlation workshop at ICTP Trieste, I ended up predicting \cite{GBPWave} p-wave superconductivity in Sr$_2$RuO$_4$, independently of Rice and Sigrist \cite{RiceSigrist}. Sr$_2$RuO$_4$ is a remarkable multi orbital system where competing interactions, a Hund coupling (and spin orbit coupling to some extent), remove a potential RVB high Tc superconductivity, and paves way for a low temperature spin triplet chilral p-wave supercondudctivity. The lost spin singlet correlations appear\cite{NeutronSRO} however, as commensurate antiferromagnetic fluctuations (similar to one seen in cuprates) in neutron scattering experiments. 

\textbf{Competing Valence Bond Order and Ultra Low Tc Superconductivity in Bi Crystal}.
Bismuth, a semimetal has a very low carrier density of about one free carrier per 10,000 Bi atoms. Inspite of such a low density and against the prediction of BCS phonon theory, crystalline Bi has been shown to exhibit an ultra low Tc $\sim$ 0.5 mK superconductivity \cite{BiTIFR} in an exciting experiment. A close look at this system by the present author shows \cite{GBBi} that Bi is well thought of as a correlated metal that hosts RVB superconductivity, via a three dimensional network of tight binding half filled quasi one dimensional chains. Chain like hybridization arises from mutually perpendicular and singly occupied 6p$_x$, 6p$_y$ and 6p$_z$ orbitals, in a reference cubic lattice. A strong electron lattice coupling stabilizes a strong and competing valence bond order, rather than a valence bond resonance. Resulting lattice dimerization (A7 structure) crashes a high Tc superconductivity. The above realization suggests ways to resurrect a lost high Tc RVB superconductivity in Bi,Sb,As and P in the A7 structure.
~~\\

\centerline{\bf{{\large 5. Five Fold Way}}}
~~\\
Inspite of the above hurdles, experimental colleagues continue to discover new high Tc superconductors
with Tc that has been on a slow rise. Based on study of cuprates and my subsequent works indicated above, I suggested a \emph{5 Fold Way to New Superconductors} \cite{5FoldWay}, a guideline for experimental search for new superconductors. Here electron correlations in relatively isolated narrow bands at the chemical potential play a central role. The five routes are: i) Copper Route (doped spin-half Mott insulator), 2) Pressure Route (Self Doped Mott Insulator in Organics), 3) Diamond Route (RVB physics in impurity band), 4) Graphene Route (Broad band 2d and intermediate correlations) and 5) Double RVB Route (Fe Arsenide family, FeSe family and self doped spin-1 Mott insulating systems).

{\bf Acknowledgement} I thank Science and Engineering Research Board (SERB), Government of India for the SERB Distinguished Fellowship.  Additional support was provided by the Perimeter Institute for Theoretical Physics. Research at Perimeter Institute is supported by the Government of Canada through the Department of Innovation, Science and Economic Development Canada and by the Province of Ontario through the Ministry of Research, Innovation and Science.

%\begin{widetext}

%\end{widetext}
\end{document}